\documentclass[prd,aps,superscriptaddress,tightenlines,nofootinbib,floatfix,preprintnumbers,11pt,longbibliography]{revtex4-1}

\usepackage[utf8]{inputenc}

\usepackage{epsfig}
\usepackage{amsfonts}
\usepackage{amsmath}
\usepackage{amssymb}
\usepackage{dsfont}

\usepackage{xspace}
\usepackage{slashed}

\usepackage{amssymb}

\usepackage{amsmath}
\usepackage{graphicx}
\usepackage{hyperref}
\usepackage[usenames]{xcolor} 

\hypersetup{colorlinks=true,allcolors=[rgb]{0.5,0.0,0.5}}

\begin{document}
\newcommand\snowmass{\begin{center}\rule[-0.2in]{\hsize}{0.01in}\\\rule{\hsize}{0.01in}\\
\vskip 0.1in Submitted to the  Proceedings of the US Community Study\\ 
on the Future of Particle Physics (Snowmass 2021)\\ 
\rule{\hsize}{0.01in}\\\rule[+0.2in]{\hsize}{0.01in} \end{center}}

\title{
Testing neutrino flavor models
}

\author{Julia  Gehrlein}
\affiliation{High Energy Theory Group, Physics Department, Brookhaven National Laboratory, Upton, NY 11973, USA}

\author{Serguey Petcov}
\affiliation{SISSA/INFN, Via Bonomea 265, 34136 Trieste, Italy}
\affiliation{Kavli IPMU (WPI), UTIAS, The University of Tokyo, Kashiwa, Chiba 277-8583, Japan}

\author{Martin Spinrath}
\affiliation{Department of Physics, National Tsing Hua University, Hsinchu, 30013, Taiwan
}
\affiliation{Center for Theory and Computation, National Tsing Hua University, Hsinchu, 30013, Taiwan}

\author{Arsenii Titov}
\affiliation{Departament de Física Teòrica, Universitat de València and IFIC, Universitat de València--CSIC, Dr.~Moliner 50, 46100 Burjassot, Spain}

\begin{abstract}
Finding a rationale behind the observed pattern of neutrino mixings has been at the focus of neutrino flavor 
model building. Many different approaches have been put forward including models based on symmetries. Among the most predictive models based on symmetries are models which predict not only the mixing parameters but also correlations between them. These mixing sum rules allow to probe flavor models in the future. In this white paper we collect the predictions for the mixing parameters from flavor models based on discrete symmetries broken to certain residual symmetries of the lepton mass matrices and from models with modular symmetries to contrast them with bounds from current and future oscillations experiments.
\end{abstract}

\snowmass
\maketitle
\section*{Executive summary}
The observation of neutrino oscillations has introduced more new parameters to the Standard Model but has also prompted a great deal of experimental and theoretical activity. Of particular theoretical interest is the question about the rationale and the origin of the mixing parameters. Approaches to the neutrino flavor puzzle include models based on symmetries, an approach motivated by the enormous success of using symmetries as the underlying guiding principle in the construction of the Standard Model. Among the most predictive flavor models are models based on symmetries which predict apart from values for the mixing parameters also correlations between them, so called mixing sum rules, which allow to test and distinguish them. These predictions in turn can provide targets for upcoming neutrino experiments and motivate the sensitivity these experiments aim to achieve in the measurement of the mixing parameters. 
We have summarized the predictions from models with mixing sum rules based on discrete symmetries broken to certain residual symmetries of the lepton mass matrices and models with modular symmetries in 
Figs.~\ref{fig:th12} and \ref{fig:delta}. Probing these flavor models is crucial to find
out whether the patterns observed in lepton mixing correspond to an underlying symmetry and will provide tremendous insights in the construction of a model beyond the Standard Model.

\newpage
\section{Introduction}
The flavor puzzle, the question of the origin of the values of the fermion masses and mixings, is one of the major open questions in particle physics. Despite the enormous success of the Standard Model (SM) we do not know yet why there are three fermion generations, and what determines the values of the fermion masses and the mixing patterns.
In fact, the vast majority of the free parameters of the SM are related to the flavor sector and describe the fermion masses and quark and lepton mixings. Finding a rationale behind the observed values of the masses and mixings is of great importance to advance our understanding of the underlying physics and will be crucial in our quest to find the ultimate model of nature.

The observation of large mixings in the lepton sector  \cite{Esteban:2020cvm}  and the existing upper bounds on the absolute neutrino mass scale \cite{Planck:2018vyg, KATRIN:2022hbd} have added a new piece to the flavor puzzle.  
The quark mixing matrix, the CKM matrix \cite{Zyla:2020zbs}, is nearly diagonal with small hierarchically ordered off-diagonal elements, whereas the
leptonic mixing matrix, the PMNS matrix, has entries of $\mathcal{O}(1)$ apart from the 1-3 entry. Even though we have not measured yet the absolute neutrino mass scale, the upper limit on the neutrino masses shows that they are at least 6 orders of magnitude lighter than the lightest charged fermion, the electron. 
Due to the prominent role of neutrinos in the flavor puzzle they might be related as well to its solution. In particular finding an explanation behind the leptonic mixing pattern has led to a plethora of flavor models in the past. 
Flavor models which fail to predict mixing parameters in agreement with experimental data are obviously not valid. However, for the remaining models the question might arise ``How do we know which flavor model is the correct one, if any?".
As we will show in the following precision measurements of the leptonic mixing parameters are fundamental to finding the correct neutrino flavor model.
While the  leptonic mixing angles have been determined with 
a rather good accuracy \cite{Esteban:2020cvm}, the Dirac CP-violating phase is largely experimentally undetermined. We will demonstrate that a precise measurement of this quantity will allow us to distinguish different flavor models from each other and therefore provide insights in not only the open question of leptonic CP violation  but also in physics beyond the SM.
 
In the following we will focus on models based on discrete flavor symmetries which lead to certain specific predictions
for the values of and/or correlations between the low-energy neutrino mixing parameters which
can be tested experimentally.

\section{Neutrino flavor models}

\subsection{Models based on symmetries}
Using symmetries as underlying guiding principle in model building has been very 
successful in the past, in fact symmetries were fundamental to construct the SM demonstrating  that nature respects certain symmetries.
Due to the success of symmetries  it is natural to expect that they might be also the clue to the solution of  the flavor puzzle.
Several classes of models have been put forward which utilize discrete symmetries, modular symmetries, or  continuous symmetries. All of these models have different predictions and phenomenology and are based on different underlying assumptions.

\subsubsection{Discrete symmetries}
Discrete non-Abelian symmetries 
have been invoked to explain the large leptonic mixing angles. As these symmetries allow for  
rotations in the flavor space by fixed (large) angles,  
 neutrino
mixing, as suggested, e.g., in \cite{Lam:2008rs}, seems to be the appropriate flavor related structure to
search for evidence of existence of an underlying flavor symmetry, and therefore for new physics.
Models based on discrete symmetries like 
 $S_3$, $S_4$, $A_4$, $T^{'}$, $A_5$ as well as the series $D_n$, $\Delta(3n^2)$, $\Delta(6n^2)$ with $n \in \mathbb{N}$ and $\Sigma$ groups (see, e.g., \cite{Altarelli:2010gt,Ishimori:2010au,King:2013eh,Feruglio:2019ybq} for
reviews and original references) have been considered.
These models predict different values for $\theta_{12}$, maximal $\theta_{23}$ but vanishing $\theta_{13}$.  Models of interest are bi-maximal (BM), tri-bimaximal (TBM),  hexagonal (HG) mixing, and models which involve the golden ration (GR)  $r=(1 +\sqrt{5})/2$ (GRA, GRB)
\begin{align}
\theta_{12}&=45^\circ ~\text{(BM)},~ \theta_{12}=\arcsin(1/\sqrt{3}) \approx 35^\circ~\text{(TBM)},~\theta_{12}=30^\circ~\text{(HG)}, \nonumber\\
\theta_{12}&=\arctan(1/r) \approx 31^\circ ~\text{(GRA)},~ \theta_{12}=\arccos(r/2) = 36^\circ ~\text{(GRB)}. 
\label{eq:symforms}
\end{align}

 The observation of non-zero $\theta_{13}$ implies that these models need to be corrected such that $\theta_{13}$, as well as $\theta_{12}$ and $\theta_{23}$, are compatible with the experimental data. A natural origin of these corrections comes from the charged lepton sector since the charged lepton mixing matrix $U_e$ contributes to the PMNS matrix as $U_{\text{PMNS}} = U^\dagger_e U_\nu$.
 In particular, models based on grand unified theories (GUT) provide a natural origin for non-diagonal charged lepton mass matrices.  In models based on $SU(5)$ $\theta_{12}^e$ is expect to be of the order of the Cabibbo angle $\theta_C$ leading to a $\theta_{13}$ in agreement with experimental data as $\theta_{13}\approx \theta_C/\sqrt{2}$ \cite{Antusch:2011qg,Marzocca:2011dh,Antusch:2009gu,Antusch:2012fb}.
In general if $U_e$ is of a 
simple form of a $U(2)$ transformation in a plane
or a product of two $U(2)$ transformations each in a plane, 
it has been shown in \cite{Petcov:2014laa,Girardi:2015vha} that a sum rule for $\cos\delta$  arises which involves the mixing angles  and $\theta_{12}$ predicted by the underlying
symmetry form of the PMNS matrix. 
An example for a sum rule is \cite{Petcov:2014laa},
$$\cos\delta = \frac{\tan\theta_{23}}{\sin2\theta_{12}\, \sin\theta_{13}} 
\left[\cos2\theta^\nu_{12} + \left(\sin^2\theta_{12} - \cos^2\theta^\nu_{12}\right)
\left(1 - \cot^2\theta_{23}\,\sin^2\theta_{13}\right)\right],$$
where $\theta_{12}$, $\theta_{13}$, $\theta_{23}$ are the measured neutrino mixing angles 
and $\delta$ is the Dirac CP-violating phase 
in the standard parametrization of the neutrino mixing 
matrix~\cite{Tanabashi:2018oca}, 
and $\theta^\nu_{12}$ is  fixed by 
the assumed underlying symmetry. 
In particular, $\theta^\nu_{12}$ can take values given in Eq.~\eqref{eq:symforms}.
Analogous sum
rules for $\cos \delta$  arise when, e.g., the TBM symmetry form of $U_\mathrm{PMNS}$ is “perturbed” on the
right by a matrix describing a $U(2)$ transformation in the 1-3 plane~\cite{Grimus:2008tt} or 2-3 plane~\cite{Albright:2008rp}, 
leading to the trimaximal mixing patterns.
Crucial for the presence of a sum rule is that less free parameters than observables are introduced. Therefore models leading to sum rules can be considered the most  economic approach to the flavor puzzle.
For systematic studies on sum rules, see \cite{Ge:2011ih,Ge:2011qn,Marzocca:2013cr,Petcov:2014laa, Girardi:2014faa, Girardi:2015zva, Girardi:2015vha,Girardi:2016zwz,Girardi:2015rwa}, and for reviews, see \cite{King:2013eh,King:2014nza,Petcov:2017ggy}.
In \cite{Petcov:2014laa, Girardi:2015vha} different mixing sum rules
have been derived, and in \cite{Girardi:2014faa,Girardi:2015zva,Girardi:2015vha}
the phenomenological consequences of these sum rules have been studied.
In \cite{Girardi:2015rwa} sum rules and predictions for $\cos \delta$ have been
obtained from 
all possible types of residual symmetries in the charged lepton and
neutrino sectors, and in~\cite{Petcov:2018snn} viability of these scenarios has been 
analyzed in light of global neutrino oscillation data. 
In sec.~\ref{sec:pheno} we will show the predictions from sum rules derived in models with certain residual symmetries for all mixing parameters.

The presence of mixing sum rules significantly increases the testability of flavor models and can guide the requirements on the target precision for the measurements of the mixing parameters.
Indeed, with the help of mixing sum rules one can forecast the
impact of measurements at future 
neutrino oscillation 
experiments~\cite{Petcov:2018snn,Antusch:2007rk,Hanlon:2013ska,Ballett:2013wya,Ballett:2014dua,Agarwalla:2017wct,Blennow:2020ncm,Blennow:2020snb} like DUNE, T2HK, ESS$\nu$SB, and JUNO.

\subsubsection{Modular symmetries}
Recently a new approach to flavor model building based on modular symmetries has been put forward \cite{Feruglio:2017spp}. In this approach modular invariance plays the role of the flavor symmetry and couplings of
the theory are modular forms of certain level $N$ and weight $k$, 
see~\cite{Feruglio:2017spp,Kobayashi:2018vbk,Penedo:2018nmg,Criado:2018thu,Kobayashi:2018scp,Novichkov:2018ovf,Novichkov:2018nkm,deAnda:2018ecu,Okada:2018yrn,Kobayashi:2018wkl,Novichkov:2018yse} for early models. 
In minimal models, the vacuum expectation value of a complex field (modulus) is the only source of flavor symmetry breaking 
and a small number of other free model parameters are present. 
Therefore, correlations between the mixing parameters arise. 
In \cite{Gehrlein:2020jnr} new sum rules in models based on modular symmetries  have been derived for the case of a fixed modulus. It has been shown that the four mixing parameters depend on two free parameters only.
An example for these relations as a function of the model parameters $\theta,~\phi$ is
 \cite{Grimus:2008tt,Petcov:2017ggy,Gehrlein:2020jnr} 
 (see also~\cite{Novichkov:2018yse})
\begin{align}
\label{eq:sr1_mix1}
 \sin^2 \theta_{12}(\theta) &= \frac{1}{3 - 2 \sin^2 \theta} \;, \\
 \label{eq:sr1_mix2}
 \sin^2 \theta_{13}(\theta) &= \frac{2}{3} \sin^2 \theta \;, \\
 \label{eq:sr1_mix3}
 \sin^2 \theta_{23}(\theta,\phi) &= \frac{1}{2} + \frac{\sin \theta_{13}(\theta) }{2} \frac{\sqrt{2 - 3    \sin^2 \theta_{13}(\theta) }}{1 - \sin^2 \theta_{13}(\theta) } \cos \phi \;, \\
 \label{eq:sr1_mix4}
 \delta(\theta,\phi) &= \arcsin\left(- \frac{\sin \phi}{\sin 2 \theta_{23}(\theta,\phi) } \right) \;.
\end{align}
The relations in other models are similar, $\theta_{12}$ and $\theta_{13}$ only depend on $\theta$ while $\theta_{23}$ and $\delta$ additionally depend on $\phi$. The exact relations however depend on the models. 
It has been pointed out first in \cite{Novichkov:2018ovf} that in the lepton 
flavor models based on modular invariance, there is a new type of 
correlation (not present in the models with traditional discrete symmetries) between the values of the neutrino masses 
(i.e., absolute neutrino mass scale) and the values of the neutrino mixing angles.
In \cite{Gehrlein:2020jnr} the correlations of $\theta_{23}$ and $\delta$ with the absolute neutrino mass scale have been studied in the cases of several analytic sum rules.
We show the predictions of the sum rules derived in models based on modular symmetries in sec.~\ref{sec:pheno}.

\subsubsection{Continuous symmetries}
Apart from discrete symmetries also  continuous symmetries have been proposed as approach to the flavor puzzle. For example models based on $
U(1), SU(2), U(2), SU(3), U(3)$ have been considered, see e.g.  \cite{Froggatt:1978nt,Barbieri:1996ae,Barbieri:1996ww,Barbieri:1997tu,Petcov:1982ya,King:2003rf}. These models are motivated by the observed hierarchies in the fermion masses for which they provide a dynamical origin and simultaneously  predict the mixing patterns. The mixing parameters are then given as functions of the Yukawa matrix elements. 
Using global, continuous symmetries which are broken by scalars (flavons) to reproduce the fermion masses and mixings leads to unobserved massless Goldstone bosons. They can be avoided by gauging the flavor symmetry as it has been done in 
 \cite{Alonso:2016onw} where the lepton flavor symmetry of the SM in the absence of lepton masses has been considered. It has been shown that the leptonic masses and mixings can be reproduced \cite{Alonso:2011yg,Alonso:2012fy,Alonso:2013nca,Alonso:2013mca}. 
 A maximal $\theta_{23}$ angle suggests that muon and tau neutrinos are maximally mixed such that a $L_\mu-L_\tau$ symmetry might be realized in nature. However, this symmetry then needs to be necessarily broken to allow for non-zero $\theta_{12},~\theta_{13}$ \cite{Fuki:2006ag} as well as the hierarchy in the 
 charged fermion masses \cite{Smirnov:2005kw}.
 
\subsection{Other approaches}

   \subsubsection{Texture zeros}
The predictivity of flavor models can be increased by a reduction of free parameters in the neutrino mass matrix. A simple way to achieve this is to
assume that some entries in the neutrino mass matrix vanish. These texture-zeros models have been thoroughly studied in the literature (see \cite{Frampton:2002yf} for early work and \cite{Mohapatra:2006gs} for a review).
Models with three-zero texture have been ruled out experimentally \cite{Zee:1980ai,Wolfenstein:1980sy,Koide:2001xy} and 
out of 15 possibilities for two-zero textures only seven   textures  lead to predictions compatible with the experimental data \cite{Frampton:2002yf} 
(for a recent analysis, see also \cite{Alcaide:2018vni}). 
These models lead to different predictions for the observable in neutrinoless double beta decay, and for $\theta_{13}$. 
However, questions about the origin of these zeros arise as well as the impact of potential corrections to the vanishing matrix elements which would spoil the predictions derived with exact texture zeros.

\subsubsection{Anarchy}

An alternative approach to the flavor puzzle not based on symmetries is anarchy \cite{Hall:1999sn,deGouvea:2012ac}. In these models 
one assumes that the elements of the neutrino mixing matrix are of comparable size without any specific underlying pattern. 
By sampling random neutrino matrices it has been shown that one expects large neutrino mixing angles. In fact,  the anarchy hypothesis is consistent with our current understanding of lepton mixing, and
the observed values for the mixing parameters are compatible (at $\sim2\sigma$ C.L.) with the ones obtained by randomly drawing a mixing
matrix from an unbiased distribution of unitary $3\times 3$ matrices. Additionally, the probability peaks at $ \sin \delta = \pm1$ \cite{Haba:2000be}  such that a measurement of maximal CP violation in the neutrino sector would be well in agreement with the anarchy hypothesis.

Anarchy does not predict correlations between different mixing parameters as their   unique probability distributions are uncorrelated. Furthermore,
anarchy only leads to probabilities for a certain measurement therefore it is challenging  to fully verify or falsify  anarchy models.

\section{Precision measurements of mixing parameters}
\label{sec:pheno}
\begin{figure}
\centering
\includegraphics[height=0.24\textwidth]{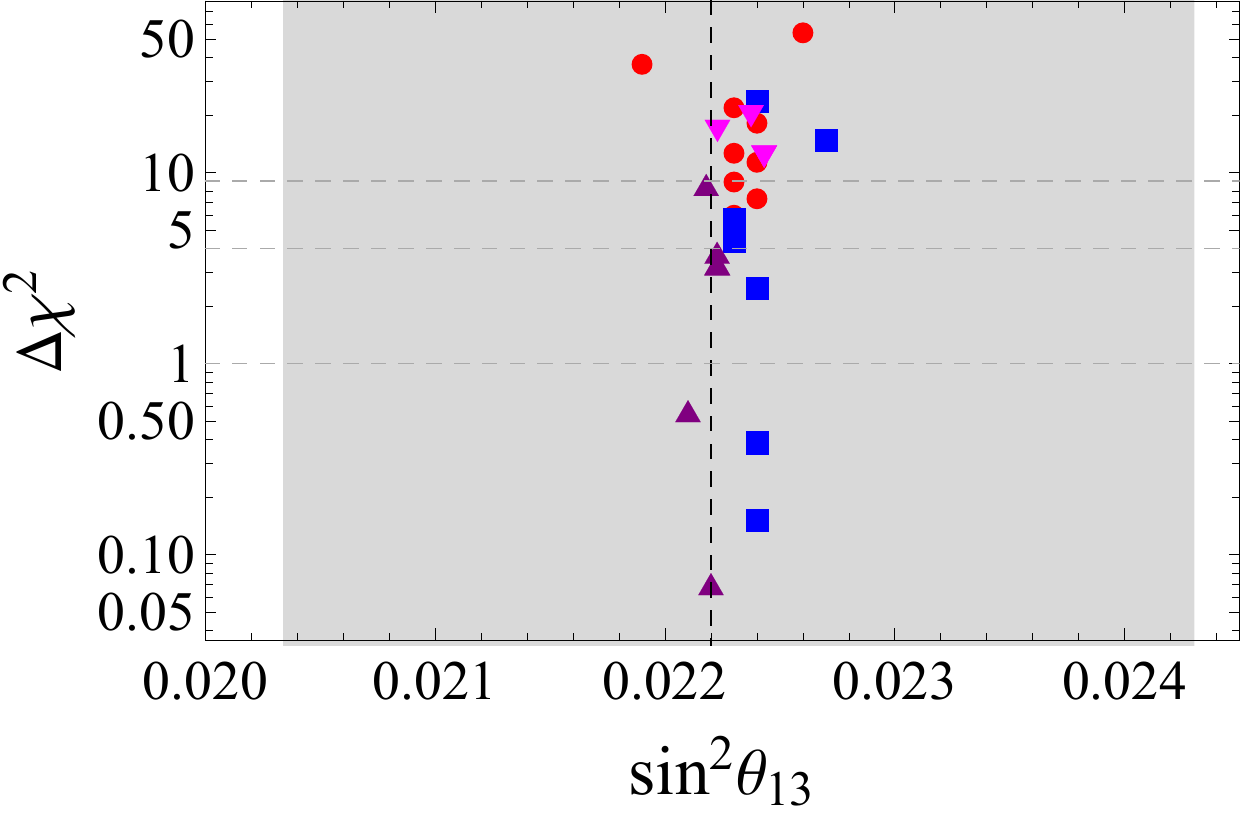}
\hfill
\includegraphics[height=0.24\textwidth]{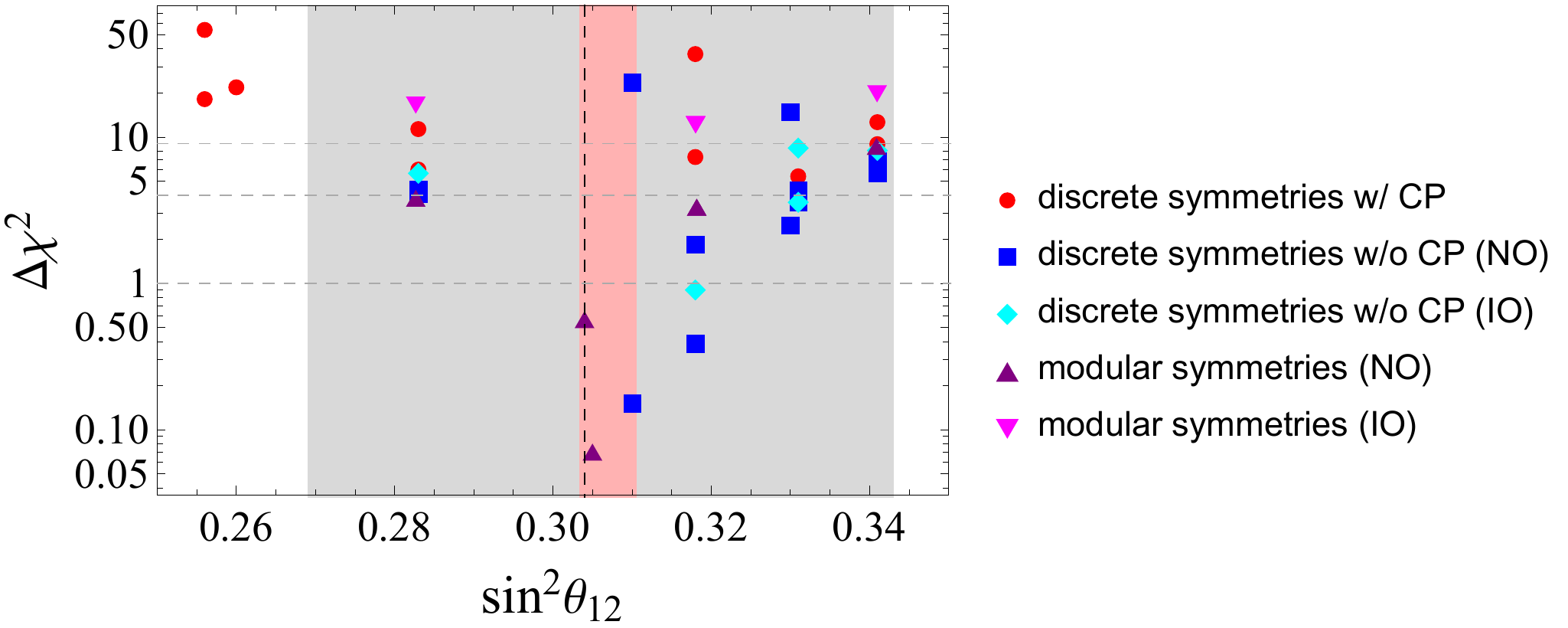}
\caption{Best-fit predictions of the models based on discrete symmetries 
broken to certain residual symmetries of the lepton mass matrices \cite{Petcov:2018snn,Blennow:2020snb}, and the models based on modular symmetries discussed in \cite{Gehrlein:2020jnr,Novichkov:2020eep,Novichkov:2021evw}. 
The gray regions are the current $3\sigma$ ranges 
for normal ordering (NO) of neutrino masses from~\cite{Esteban:2020cvm}
(the ranges 
for inverted ordering (IO) are very similar). The  dashed line is the current best fit value. The red region is the prospective $3\sigma$ range 
after 6 years of JUNO running~\cite{JUNO:2015zny}. 
The assumed  future best fit value is $\sin^2\theta_{12}=0.307$. 
The sensitivity of DUNE to $\theta_{13}$  after 15 years of running~\cite{DUNE:2020ypp} will not improve current bounds.
}
\label{fig:th12}
\end{figure}

\begin{figure}
\centering
\includegraphics[height=0.24\textwidth]{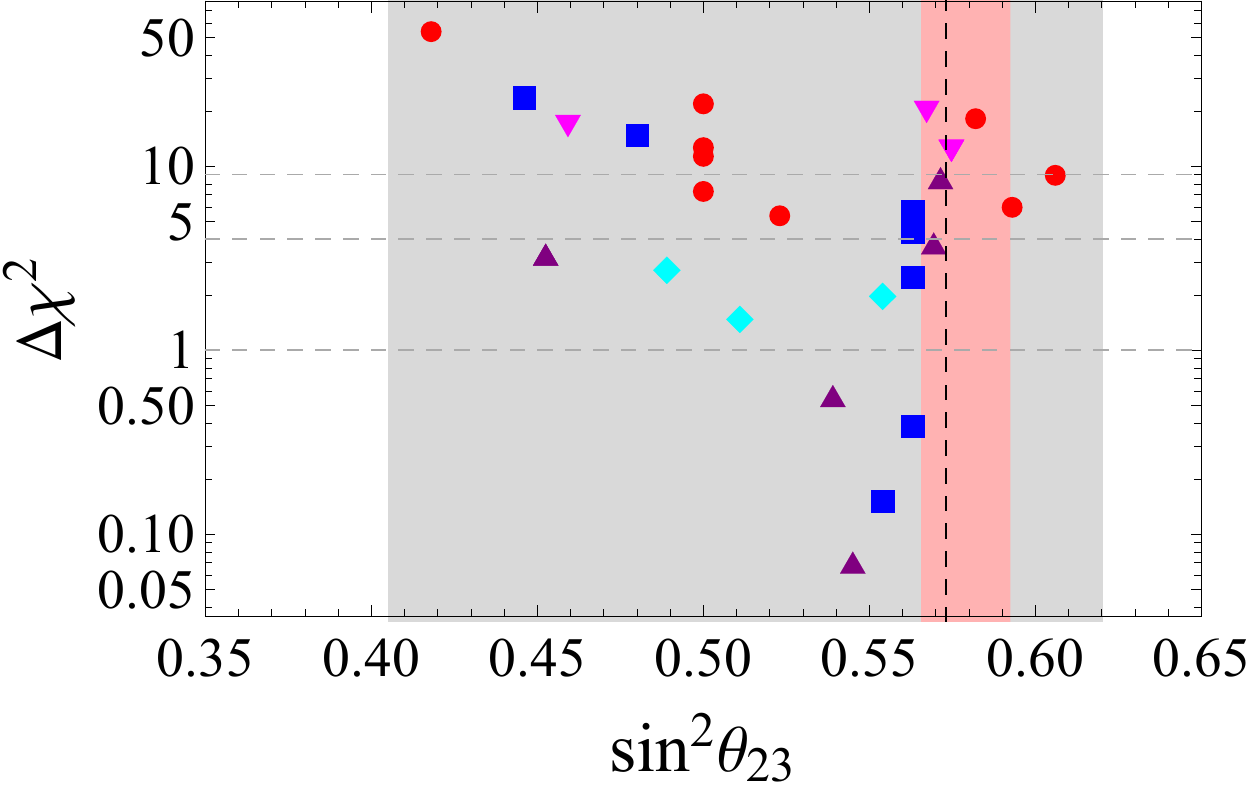}
\hfill
\includegraphics[height=0.24\textwidth]{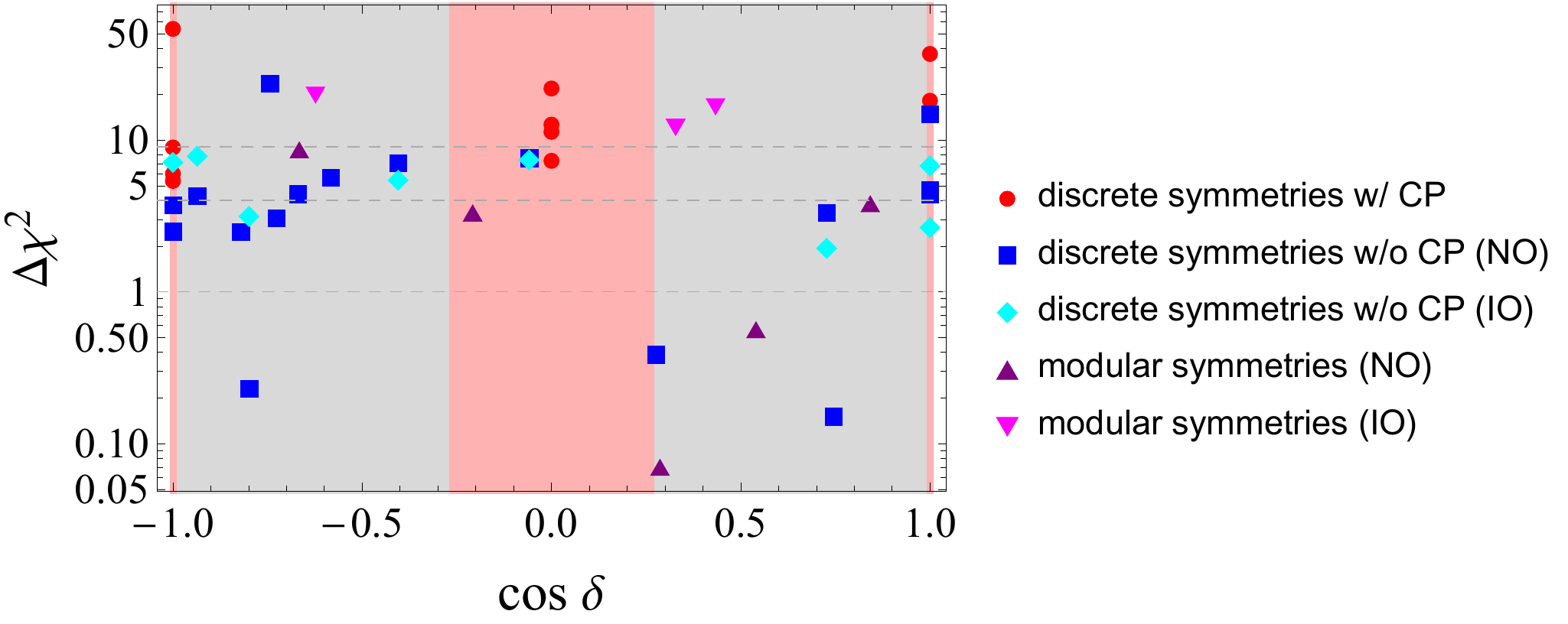}
\caption{Best-fit predictions of the models based on discrete symmetries
broken to certain residual symmetries of the lepton mass matrices \cite{Petcov:2018snn,Blennow:2020snb}, and the models based on modular symmetries discussed in \cite{Gehrlein:2020jnr,Novichkov:2020eep,Novichkov:2021evw}.
The gray regions are the current $3\sigma$ ranges for NO from \cite{Esteban:2020cvm} (the ranges for IO are very similar). The  dashed line is the current best fit value for NO. The red regions are the prospective $3\sigma$ ranges after 15 years  of DUNE running~\cite{DUNE:2020ypp}. The assumed future best fit value is $\sin^2\theta_{23}=0.58$. The current $3\sigma$ range for $\cos\delta$ extends over  the whole parameter range. 
For the prospective $3\sigma$ ranges
we show the sensitivities for assumed true values $\delta=0$ and $\delta=-\pi/2$.
}
\label{fig:delta}
\end{figure}

Predictions from flavor models based on discrete and modular symmetries provide excellent targets for neutrino oscillation experiments. In addition to predicting the mixing parameters, they also predict correlations between them enhancing therefore their testability and providing an additional handle to disentangle different models. We show in 
Figs.~\ref{fig:th12} and \ref{fig:delta}
the best fit predictions for the mixing parameters in the three classes of flavor models:
\begin{itemize}
 \item Models based on the non-Abelian discrete symmetries $A_4$, $S_4$ and $A_5$ combined with a generalized CP symmetry~\cite{Feruglio:2012cw,Holthausen:2012dk} 
and broken to $G_e > Z_2$ and $G_\nu = Z_2 \times \mathrm{CP}$ residual symmetries
in the charged lepton and neutrino sectors, respectively. 
The corresponding mixing patterns are characterized by one free angle parameter. 
See~\cite{Blennow:2020snb} for a summary of such scenarios and original references.
 \item Models based on the non-Abelian discrete symmetries $A_4$, $S_4$ and $A_5$ broken to $G_e~(G_\nu) > Z_2$ and $G_\nu~(G_e) = Z_2$~\cite{Girardi:2015rwa,Petcov:2018snn}. 
The corresponding mixing patterns are determined by two free parameters~---~an 
angle and a phase.
 \item Models with modular $A_4$, $S_4$ and $A_5$ symmetries~\cite{Novichkov:2018yse,Novichkov:2018ovf,Novichkov:2018nkm,King:2019vhv} 
 leading to the sum rules derived in \cite{Gehrlein:2020jnr}, 
as well as the models based on the double cover of modular $S_4$~\cite{Novichkov:2020eep,Novichkov:2021evw} 
combined with a generalized CP symmetry~\cite{Novichkov:2019sqv}. 
In the cases when analytical sum rules 
can be derived, the resulting mixing patterns depend on two free parameters, 
cf.~Eqs.~\eqref{eq:sr1_mix1}--\eqref{eq:sr1_mix4}. 

\end{itemize}
The corresponding total $\Delta\chi^2$ is calculated using 
the results of the global analysis of neutrino oscillation data performed in \cite{Esteban:2020cvm}.
The predictions for the mixing parameters do not depend on the neutrino mass ordering. However, as the model parameters have been fitted to the global neutrino data, the $\Delta\chi^2$ of the predictions changes.
Note that since $\sin^2\theta_{13}$ is determined with a relatively high precision, it drives the value of the free angle parameter when fitting the models 
to the data, cf.~Eq.~\eqref{eq:sr1_mix2}. This is why in all the cases considered, 
a value of $\sin^2\theta_{13}$ lying very close to its experimental best-fit value is realized.
We compare the model predictions with the current constraints \cite{Esteban:2020cvm} and projections from upcoming neutrino experiments like DUNE~\cite{DUNE:2020ypp}  and JUNO~\cite{JUNO:2015zny}.

We see that a precise measurement of the mixing parameters will be crucial to probe and disentangle flavor models. However, the required sensitivity to distinguish between different models  depends on the true value of the parameter, 
as there are classes of models which have very similar predictions such that an isolated measurement of one angle cannot distinguish these models. Therefore the correlations between the mixing parameters should be probed, these can be even more powerful than isolated measurements of the angles \cite{Blennow:2020snb, Blennow:2020snb, Gehrlein:2020jnr}. 
In \cite{Agarwalla:2017wct} it has been shown that the combined data of DUNE and T2HK can distinguish some models based on discrete symmetries from each other at the $3\sigma$ level.
The predicted values for $\theta_{12}$ deviate from the current best fit value whereas the predictions for $\theta_{13}$ are rather close to the current best fit.   If the real  value of $\theta_{12}$ is indeed close to the current best fit the majority of the flavor models discussed here will be disfavored. Therefore it will be imperative that JUNO achieves the envisioned precision in the determination of $\theta_{12}$.
The model predictions for $\cos\delta$ and $\theta_{23}$ are rather distributed within the current $3\sigma$ ranges. Thus already with moderate precision in $\delta$ and $\theta_{23}$ many models can be probed and distinguished.
While T2HK~\cite{Hyper-Kamiokande:2018ofw} will have only  sensitivity to $|\cos(\delta)|$, DUNE will have access to the sign. This could become interesting if the true value of  $\delta$ is not near $\pm\pi/2$ as predicted by many flavor models.

In these studies it was assumed that the sum rule is exactly
realized at low energy.  However, as every quantity in quantum field theory,
the mixing parameters get affected by renormalization group running which has
been recently updated and studied systematically in \cite{Gehrlein:2016fms,Zhang:2016djh,Zhang:2016png}.
Those corrections can be particularly important in GUT scenarios where these sum rules can 
occur as well~\cite{Antusch:2011qg,Marzocca:2011dh,Antusch:2012fb}.
Future precision measurements of the mixing parameters need to be carefully compared to flavor model predictions taking possible corrections to these relations into account.

\section{Conclusions}
We are entering the precision era of neutrino physics.
But the question arises ``How precise do we want to measure the neutrino parameters?". Flavor models can provide a compelling answer to this question. 
Among the most predictive flavor models are models based on discrete or modular symmetries as they predict  concrete values for the mixing parameters as well as correlations between them. These sum rules can be used to distinguish different flavor models. In Figs.~\ref{fig:th12} and \ref{fig:delta} we have compiled the predictions of these models and demonstrated that precision measurements of all mixing parameters are crucial to thoroughly test flavor models. Therefore it is of utmost importance that upcoming neutrino oscillation experiments like DUNE, T2HK, and JUNO reach their envisioned precision. Doing so will guide us in our quest to unveil new physics  beyond the SM by testing the existence of additional symmetries.
 In the past  precision measurements of the mixing angles were already fundamental in testing flavor models as  the measurement of 
non-zero $\theta_{13}$ with small uncertainties excluded many flavor models \cite{Albright:2006cw}. This demonstrates the power of precision measurement in neutrino physics in our path to new physics models.

In addition to precise measurements of the mixing parameters,
models based on continuous symmetries  lead to additional predictions like  new degrees of freedom. In particular the models based on continuous gauged symmetries (see for example \cite{Heeck:2011wj,Alonso:2016onw}) predict new gauge bosons with a certain flavor structure and these models can  therefore  also be probed at the intensity and energy frontiers. 
While we focused here on the predictions of flavor models for the mixing parameters, models where the light neutrino masses depend on two free parameters only \cite{Gehrlein:2017ryu}  predict relations between the light neutrino masses including the Majorana phases~\cite{Bilenky:1980cx}. These  mass sum rules can be probed at experiments sensitive to the absolute neutrino mass scale, the neutrino mass ordering, and with neutrinoless double beta decay experiments \cite{Barry:2010yk,Dorame:2011eb,King:2013psa,Gehrlein:2016wlc, Gehrlein:2015ena}.
Therefore, in addition to precision measurements of mixing angles also an observation of neutrinoless double beta decay can provide insights into flavor symmetries. 
Together with the predictions for the Majorana phases (see, e.g.,~\cite{Girardi:2016zwz}), 
flavor models predicting the Dirac phase could be 
 related to the production of the matter asymmetry of the Universe~\cite{Bertuzzo:2009im,Chen:2016ptr,Hagedorn:2016lva} 
by providing the required amount of CP violation 
(see, e.g.,~\cite{Pascoli:2006ci,Moffat:2018smo,Granelli:2021fyc}).

Flavor models provide a rich phenomenology and represent an ideal target for upcoming experiments. Future detailed studies of sum rules both from the theoretical as well as from the experimental side are hence a great opportunity for model builders, phenomenologists and experimentalists to gain more insights into the mysteries hidden in the neutrino sector.

\section*{Acknowledgements}
J.G. acknowledges support from the US Department
of Energy under Grant Contract DE-SC0012704. 
The work of S.T.P. was supported in part by  
 the European Union's Horizon 2020 research and innovation programme under the Marie Skłodowska-Curie grant agreement No.~860881-HIDDeN,
 the INFN program on Theoretical Astroparticle Physics 
and by the  World Premier International Research Center
Initiative (WPI Initiative, MEXT), Japan. 
M.S. is supported by the Ministry of Science and Technology (MOST) of Taiwan 
under grant numbers 
MOST 107-2112-M-007-031-MY3 and MOST 110-2112-M-007-018.
The work of A.T. is supported by the “Generalitat Valenciana” under grant PROMETEO/2019/087,
as well as by the FEDER/MCIyU-AEI grant FPA2017-84543-P and the MICINN-AEI grant PID2020-113334GB-I00.

\bibliographystyle{apsrev}

\bibliography{sr}

\end{document}